\documentstyle[twoside, epsf]{article}
\input{epsf.sty}

\input ibvs2.sty

\begin{document}

\IBVShead{5xxx}{00 Month 200x}

\IBVStitle{Limits on Transit Timing Variations in HAT-P-6 and WASP-1}

\IBVSauth{SZAB\'O, Gy.M.$^{1,2}$, HAJA, O.$^2$, SZATM\'ARY, K.$^2$, 
P\'al, A.$^{1,3}$, Kiss, L. L.$^1$}

\IBVSinst{Konkoly Observatory of the Hungarian Academy of Sciences, P.O. Box 67, H-1525 Budapest,
Hungary; e-mail:szgy@konkoly.hu}
\IBVSinst{Dept. Experimental Physics. \& Astronomical Obs., Univ. of Szeged,
6720 Szeged D\'om t\'er 
9, Hungary}
\IBVSinst{ Department of Astronomy, Lor\'and E\"otv\"os University, P\'azm\'any 
P. st. 1/A, H-1117 Budapest, Hungary}

\SIMBADobjAlias{ HAT-P-6,WASP-1}{}
\IBVStyp{HADS}
\IBVSkey{analysis}
\IBVSabs{We detected the transit of HAT-P-6b at 
HJD 2454698.3908$\pm$0.0011 and that of WASP-1b
at HJD 2454774.3448$\pm$0.0023.}
\IBVSabs{The updated orbital periods of the planets are
3.852992$\pm$0.000005 days (HAT-P-6b) and 2.519970$\pm$0.000003 days
(WASP-1b).}
\IBVSabs{There is no indication of any departures from constant orbital period
in these systems.}

\begintext

 The study of Transit Timing Variations 
(TTV, e.g. D\'\i{}az et al. 2008, Sozzetti et al. 2009) is important because it
may reveal the effect of other perturbing
planets in the exoplanetary systems (Steffen and Agol, 2005), 
or moons of the transiting exoplanet (e.g. Szab\'o{} et al. 2006, 
Simon et al. 2007, Kipping et al. 2009ab).
We present new transit times and Transit Timing Variation analysis
of two exoplanets, HAT-P-6b and WASP-1b.

Time series were taken at two different sites. On 19/20 August, 
2008, 
HAT-P-6 was observed with the 0.6~m Schmidt
telescope of the Konkoly Observatory, Piszk\'estet\H{o} mountain station. The 
integration time was 15 s through Johnson R filter. 
On 3/4 November, 2008, we observed WASP-1 with the 0.4 m 
telescope of the Szeged Observatory, equipped with an ST-7E CCD camera.
The integration time was 30 s through Johnson I filter.

The data were analysed with aperture photometry in IRAF, with an ensemble 
of comparison stars. Stellar magnitudes
were obtained with multiple apertures. The optimal aperture size was determined
with minimizing the $rms$ scatter of the residuals. In both cases, the aperture radius 
was 4 pixels, corresponding to
4 arc seconds with the 0.6 m Schmidt and 5.3 arc seconds with the 0.4 m
Newtonian. 
The scatter of the raw light curves is $\approx \pm$0.005 mag for HAT-P-6 and $\approx \pm$0.008 mag 
for WASP-1. After calculating 3-minute averages, these values are reduced to 
$\approx \pm$0.0025 mag (HAT-P-6) and
$\approx \pm$0.004 (WASP-1). 

Times of minima were determined by fitting a model light curve. To reduce the
degree of freedom of fitting, the shape of the model was not adjusted;
we used previously published parameters.
The model was shifted in time, minimising the $rms$ scatter of the measurements.

The log of observations is summarised in Table 1, 
light curves and $TTV$ diagrams are shown in Fig 1.

\begin{table}[h]
\begin{center}
\begin{tabular}{cccccc}
Planet & Date   & HJD &      duration & number  & transit time   \\
      & & (first point) & (hour) &  of points & HJD$-$2450000\\
\hline
HAT-P-6b & 2008.08.19 & 2454698.30    &  4.5    & 647 & 4698.3908$\pm$0.0011\\
WASP-1b & 2008.11.04 & 2454774.22   &   5.2    & 351 & 4774.3448$\pm$0.0023\\
\hline       
\end{tabular}
\caption{The log of observations}
\end{center}
\end{table}

\IBVSfig{12cm}{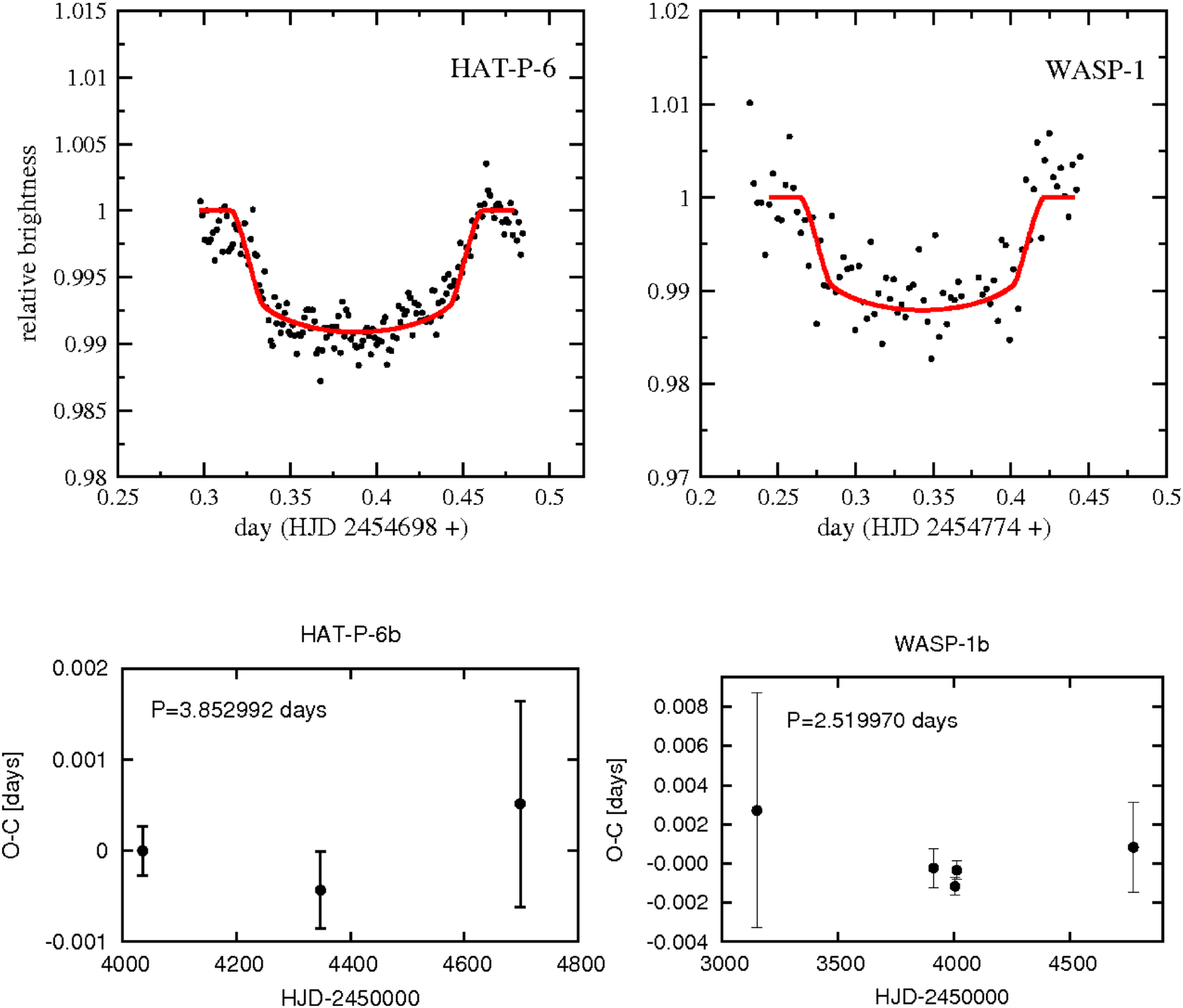}{Light curves of the observed transits and fitted
models (top panels) 
and $O-C$ diagrams of the exoplanet systems (bottom panels).}

\bigbreak

{\bf Notes on individual exoplanets:}

\bigbreak

{\bf HAT-P-6b} is a hot Jupiter known for its very low density.
One time of mid-transit at HJD
2454035.67575 $\pm$ 0.00028
was published by Noyes et al (2008), who determined a period of 
3.852985 $\pm$ 0.000005 days. 
They also published a second light curve starting from 2454347.7.
We have re-fitted the publicly available photometry 
simulataneously with our new data by assuming the same transit
geometry parameters (duration, depth, impact parameter) 
but independent transit times. 
The resulted transit times are 2454035.67571$\pm$0.00027, 
2454347.76763$\pm$0.00042 and 2454698.3908$\pm$0.0011.
The data do not indicate
any departure from constant orbital period.

{\bf WASP-1b:} is supposed to be a hot Jupiter with
metal-rich atmosphere, little or no core, and its age is less than
1.5 Gyr. (Cameron et al. 2007, Stempels et al. 2007). Two transits at
2453912.514 $\pm$ 0.001 and
2454005.75196 $\pm$ 0.00045 
were published by Charbonneau et al. (2007), who adopted a period of 
2.51997 days. Shporer et al. (2007) published a transit 
time at 2454013.31269
$\pm$ 0.00047 and determined a period of 2.519961$\pm$0.000018.
Cameron et al. (2007) have also published 
measurements from 2004 with a pre-discovery transit
at 2453151.486 $\pm$ 0.006 (the numerical value is available at exoplanet.eu);
the resulting period was 2.51995 $\pm$ 0.00001 days.

We measured a transit at HJD 2454774.3448$\pm$0.0023
and determined a new period of 2.519970$\pm$0.000003 days.
All transit times are compatible with the updated ephemeris. 
The pre-discovery transit time 
is well off
a linear fit, but the large error bar preclude any conclusion on 
TTV.
By neglecting this first point, the best-fitting period is 2.519973$\pm$0.000003 days and
the earliest point is a significant outlier. At this moment
it is unclear whether there is period change in the system, hence
further monitoring is necessary.

{\bf Acknowledgements}
The research was supported by the ``Lend\"ulet'' Program of the Hungarian Academy
of Sciences, 
Hungarian OTKA Grant K76816 and the ``Bolyai J\'anos''
postdoctoral fellowship of the Hungarian Academy of Sciences.

\references

Cameron, A. C., Bouchy, F., H\'ebrard, G., et al., 2007, {\it MNRAS}, {\bf 375}, 951

Cameron, A. C., Pollacco, D., Hellier, C., et al., 2009, IAUS, 253, 29 

Charbonneau, D., Winn, J. N., Mark, E. E., et al., 2007, {\it ApJ}, {\bf 658}, 1322

D\'\i{}az, R. F., Rojo, P., Melita, M., et al., 2008, {\it ApJ}, {\bf 682}, 49

Kipping, D. M., 2009a, {\it MNRAS}, {\bf 392}, 181

Kipping, D. M., 2009b, {\it MNRAS}, {\bf 396}, 1797

Noyes, R. W., Bakos, G. \'A., Torres, G., et al., 2008, {\it ApJ}, {\bf 673}, 79

Shporer, A., Tamuz, O., Zucker, S., Mazeh T., 2007, {\it MNRAS}, {\bf 376}, 1296 

Simon, A., Szatm\'ary, K., Szab\'o, Gy. M.,  2007, {\it A\&{}A}, {\bf 470}, 727

Sozzetti, A., 2009, {\it ApJ}, {\bf 691}, 1145

Steffen, J. H., Agol, E., 2005, {\it MNRAS}, {\bf 364}, 96

Stempels, H. C., Collier Cameron, A., Hebb, L., et al., 2007, {\it MNRAS}, {\bf 379}, 773

Szab\'o, Gy. M., Szatm\'ary, K., Div\'eki, Zs., Simon, A., 2006, {\it A\&{}A}, {\bf 450}, 395

\endreferences



\end{document}